# Towards a Computational Framework for Automated Discovery and Modeling of Biological Rhythms from Wearable Data Streams

Runze Yan[1]([✉]) and Afsaneh Doryab[1]

University of Virginia, Charlottesville VA 22903, USA,
ry4jr@virginia.edu

Abstract. Modeling biological rhythms helps understand the complex principles behind the physical and psychological abnormalities of human bodies, to plan life schedules, and avoid persisting fatigue and mood and sleep alterations due to the desynchronization of those rhythms. The first step in modeling biological rhythms is to identify their characteristics, such as cyclic periods, phase, and amplitude. However, human rhythms are susceptible to external events, which cause irregular fluctuations in waveforms and affect the characterization of each rhythm. In this paper, we present our exploratory work towards developing a computational framework for automated discovery and modeling of human rhythms. We first identify cyclic periods in time series data using three different methods and test their performance on both synthetic data and real fine-grained biological data. We observe consistent periods are detected by all three methods. We then model inner cycles within each period through identifying change points to observe fluctuations in biological data that may inform the impact of external events on human rhythms. The results provide initial insights into the design of a computational framework for discovering and modeling human rhythms.

## 1 Introduction

Biological rhythms created through periodic changes in the physiological functions and living habits of many organisms are managed by biological clocks [1]. Similar to other creatures, there is evident periodicity in various physiological and psychological activities of human bodies, such as heartbeat, electroencephalogram (EEG), and female menstruation [2,3]. Biological rhythms, however, are affected by external events such as social obligations and work schedules. Persistent disruption of these cycles may result in physical and mental health problems such as cardiovascular disease [4] and depression [5].

    Understanding and modeling human rhythms can have profound impacts on the health and well-being of individuals. The advancements of technology allow for

longitudinal tracking of physiological and behavioral data using consumer-level devices outside of the laboratory. As such, computational systems that are aware of human rhythms can help people in planning their life around their rhythms to achieve better health and wellness.

The first step in understanding and modeling biological rhythms is to identify their characteristics, such as cyclic periods, phases, and amplitude. However, so far, most studies in modeling rhythms have focused on circadian rhythms that were introduced by Halberg in the 1950s to denote the 24 hours of sleep-wake cycle [6]. Since then, numerous procedures for the numerical analysis of circadian rhythms have been proposed (See, e.g., Refinetti et al. for a summary [7]). Zielinski et al. (2014) developed an online system that incorporated six algorithms to make an accurate estimate of the underlying period of circadian data [8]. In the medical field, modeling of disruption in biobehavioral rhythms in cancer patients was used to predict readmission risk after pancreatic surgery [9]. In all of these studies, the assumption has been that an underlying 24-hour cycle in data exists.

In this paper, we aim to advance the computational modeling of human rhythms by 1) automatically discovering all possible periods of a human time series data and 2) building a hierarchical model to further identify the inner cycles of each period and thereby observe possible fluctuations and disruptions in biological rhythms resulted from external stimulus. We will first apply three well-known but fundamentally different methods, namely Fast Fourier Transform (FFT), Chi-squared based periodogram, and Cyclic Hidden Markov Models (CyHMMs) to detect periods in three datasets. The first dataset is created artificially with known periods to test how well those three methods can detect periods in the data. The methods are then tested on a real-world dataset containing heart rate and temperature data over 70 days as well as a dataset with data collected from the E4 wearable device that consists of 16 days of fine-grained heart rate (HR), heart rate variability (BVP), skin temperature, and galvanic skin response (EDA). In the next step, we explore the inner variability of physiological data in each cycle via adopting the Automatic Non-stationary Oscillatory Modelling (AutoNOM) method that can model non-stationary time series to identify change points and achieve piecewise fitting simultaneously [10]. Our framework can identify the known periods in the synthetic dataset with small errors. In the two real-world datasets, HR, skin temperature, BVP, and EDA are found to have a 24-hour circadian cycle and detected changing points gathered around 6:40 am and 8:00 pm. Compared with current approaches, our framework can model rhythms using different methods to provide insights into commonality and differences between those models. It can also inspect the signal series for fluctuations and changes that are essential to human health and wellbeing. Overall, our research makes the following contributions:

- We explore the feasibility of developing a computational framework for discovering and modeling human rhythms from multiple sensor data streams. We investigate the requrements of such framework to model the overall rhythms and to detect the instantaneous changes in the rhythmic data.
- We evaluate different periodicity detection algorithms on one synthetic dataset and two real-world physiological datasets and describe the application of changing point detection technology to rhythm modeling.

In the following sections, we describe our approach and discuss the experimental results.

## 2   Related Work

Several studies have shown the impact of biological and behavioral rhythms on health and wellbeing [11, 12]. Abdullah et al. demonstrated the circadian rhythm regularity in bipolar disorder using data collected from smartphones [13]. Murnane et al. built the connection between circadian rhythm models and phone usage and pointed out overloaded phone usage in the late evening as a cause of low-quality sleep [14]. The aforementioned studies have assumed the regularity of circadian rhythms and tried to show the impact of people's behavior (e.g., phone usage) on the sleep-wake cycle. Our research goes beyond assuming known periodicity in data and uses computational modeling to discover possible cycles (more or less than 24 hours).

Among the periodicity detection methods, Fast Fourier Transform (FFT) is the most commonly used approach. FFT converts a time function into a frequency function. The dominant frequency in the frequency function is selected as the period of rhythm [15]. Lomb-Scargle periodogram is another algorithm for cycle discovery based on Fourier analysis and is mainly used for unequal distance data [16]. Saner et al. assessed blood pressure and heart rate with Fourier analysis and found that cardiovascular rhythmicity is related to obesity in children [17]. Chi-square periodogram uses Chi-square statistics, has been applied to model the period of lobsters' circadian rhythms to enhance fishing efficiency [18]. Chi-square periodogram calculates the variance between-period and within-period, and use $\chi^2$ distribution to evaluate the significant level. Vukolic et al. applied Chi-squared to model the circadian rhythms of blood pressure and heart rate of mice carrying the muted gene Per2, and found the circadian clock gene Per2 control cardiovascular rhythms [19]. Cyclic Hidden Markov Model (CyHMM) has also been used for modeling cyclic time series data. CyHMM outputs the period's length by inferring the cyclic latent states of input time series [3]. Pierson et al. applied the cyHMM to sleep time, steps, and calories burned and found that these features showed a weekly cycle. All these statistical models could output the general periods of time series, but they all lose the detailed information within each period at the same time and do not capture the variation in the shape of sequence in each cycle [3].

Once the period is known, cyclic functions can be applied to the time series data to model the rhythmicity. Cosinor is a standard method to model the amplitude and phase of rhythms when the period is used as input. It uses a linear combination of cosine curves to fit time series data using least squares regression [20]. Doryab et al. used Cosinor to extract rhythm features from cancer patients and to predict the readmission probability using these features [9]. However, cosinor cannot fit the non-stationary time series well. Changing point detection (CPD) helps model the time series within one period; it can split the non-stationary time series into several stationary series pieces. It can also identify the location of detected changing points. CPD has been used in monitoring medical conditions. For example, applying CPD to heart rate (HR), electrocardiogram (ECG), and electroencephalogram (EEG) has helped better diagnosis of heart disease and understand brain activity [10,21–24]. CPD has also been applied to human activity recognition using data from smart home and mobile devices. The changing points in this context represent the transition of human activity [25–27]. Selection criteria differ among CPD methods, and some methods are sensitive to the changes of amplitude in the mean, variance, correlation, and spectral density. The cumulative sum (CUSUM) is the most familiar CPD algorithm. CUSUM tracks the shift of local mean, and if the decrease or increase of the mean exceeds the threshold, one change will be identified [28].

For modeling the variations in physiological data during each cycle, we adopt the Automatic Non-stationary Oscillatory Modelling (AutoNOM) to model non-stationary time series with a known period. AutoNOM identifies change points in each cycle and achieves piecewise fitting [10] using sinusoidal regression models simultaneously. We prefer the CPD technology used in the AutoNOM because it is more sensitive to the change of frequencies of time series, and the AutoNOM can find the best sinusoidal equations to fit the data in each segment [10]. The following section describes these methods in detail.

## 3    Methods

The analysis procedure of physiological data in this paper can be divided into three steps, shown in Figure 1. In the first step, the raw data is processed, cleaned, and missing values are imputed. In the second step, the period detection methods are applied to each time series dataset to infer possible significant periods ($p < 0.01$). We choose three different algorithms, including the FFT [15] based on the Fourier Transformation, the Chi-squared periodogram [29] based on the chi-square statistic, and the CyHMMs [3] based on the state transition. From this step, we choose common periods selected by all three methods to be used for the final step in which the data in each period is modeled via the AutoNOM method for detecting inner cycles and estimating their rhythmic characteristics. Once a dataset is determined to exhibit rhythmicity, we can further extract the characteristics of the rhythm, including MESOR (the average value around the variable oscillation), amplitude (half the difference between the peak and trough of the wave), and phase (the time at which the peak of rhythm occurs). These features are shown in Figure 2.

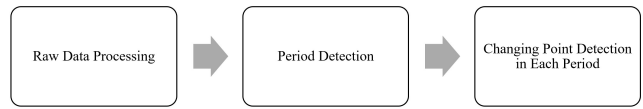

Fig.1: Analysis pipeline for the physiological data

### 3.1    Periodicity Detection

Fast Fourier Transform (FFT) Fast Fourier transform is an algorithm that converts a signal from the time domain to the frequency domain [15]. In this way, a periodic time series can be expressed by the sum of its frequency components. The Fourier periodogram obtained by the FFT encodes the spectral energy at a given frequency, and the dominant frequency is the component with maximal frequency. The dominant period is the reverse of the dominant frequency.

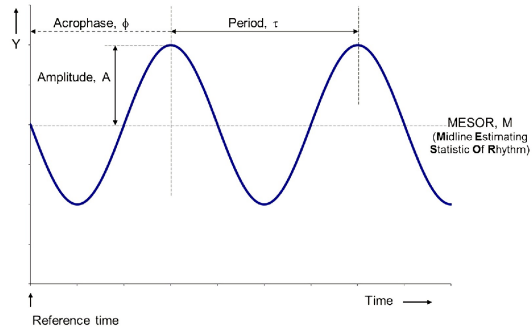

Fig.2: Rhythmic parameters to depict a periodic series [30]

Chi-squared Periodogram The Chi-squared Periodogram was developed from the Enright Periodogram [31]. The Enright periodogram is based on the principle that the variances of different segments of the time series are arranged in periodic order sequentially. This process repeatedly divides the long data stream into different periods and calculates a variability index for each period. For the significance test of each hypothesis period, the Enright periodogram uses the F statistic to compare the between-class and within-class variabilities, to test the null hypothesis of the equal class mean. The Chi-square periodogram uses the $\chi^2$ distribution instead of F distribution. Sokolove and Bushell [32] proposed the index $Q_P$ to calculate the significance of different frequencies in time-series data. The stronger is the rhythmicity in a data set, the higher is the value of $Q_P$. For a dataset with $N$ values (i.e., $X_i$ for $i = 1$ to $N$), which can be broken down into $K$ sections of period $P$, the formula of $Q_P$ could be defined as follows [33]:

$$Q_P = \frac{KN \sum_{h=1}^{P}(M_h - M)^2}{\sum_{i=1}^{N}(X_i - M)^2}$$

(1)

where $M_h$ is the mean of $P$ values under each time unit of the period length, and M is the mean of all N values.

Cyclic Hidden Markov Models Cyclic Hidden Markov Models (CyHMMs) are a special kind of Hidden Markov Models (HMMs) [34] for detecting and modeling cyclic patterns. The input time series will be treated as the observation sequence, and a set of cyclic latent states will be inferred for the observation sequence. The period of latent states will be returned as the period of input time series. Although having the same primary structure as the standard HMMs, CyHMMs differ from HMMs as they do not allow the random transition between hidden states. They require that the transition between hidden states follows a specific order, and the next state of the final state is the starting state, thus forming a closed-loop link to reflect the periodicity. In the CyHMMs, the time spent at a particular stage follows the Poisson distribution, while the standard HMMs have a Geometric distribution [3].

3.2    Changing Point Detection in Cycles

As mentioned previously, once periodicity is detected in the time series data, we want to explore the variation within each period to understand the potential impact of external factors or fluctuations in each cycle. To choose a method for analyzing each time series, we explore stationary and non-stationary time series methods. The theory and methods for analyzing the stationary time series are relatively well-developed. For example, the cosinor approach [20] fits data by a combination of cosine curves with or without polynomial terms using the least-squares method. However, the cosinor does not apply to non-stationary data, and when the waveform approximates squares, the working performance is not excellent. Since most physiological signals are non-stationary, we need to consider methods suitable for non-stationary time series [35].

As mentioned in the Related Work section, we use AutoNOM to identify change points in each cycle using sinusoidal regression models. The model can be divided into two sub-models: the segment model and the change point model. The rhythm time series $y$ is segmented by $k$ unknown change points by analyzing the frequency change over time. If the frequency on both sides of a time point has a sudden change in the frequency domain, this time point will be regarded as a substation. In each time segment, AutoNOM measures the frequency $\omega$, amplitude $\beta$, and phase of the segment $\sigma$. $\omega$, $\beta$, and $\sigma$ are multidimensional vectors, and their dimension numbers are unknown. Before using AutoNOM, we need to set the maximum number of change-points $k_{max}$ and the maximum number of frequencies on each segment $m_{max}$. These threshold setting will transform the problem of modeling rhythm series into a finite state problem. The AutoNOM method will then select an optimal model from the model state space, which is composed of models with a different number of change points and frequencies. The optimal model is the model with the maximum posterior probability, which is calculated as follows [10]:

$$\pi(k,m_k,s_k,\theta_k|y) = \pi(k|y)\pi(m_k|k,y)\pi(s_k|m_k,k,y)$$
$$\pi(\theta_k|s_k,k,m_k,y),$$

(2)

where $k$ is the number of unknown change-points, $m_k$ is the number of frequency components in each time series segment, $s_k$ is the location of change-points, and $\theta_k$ is a three-dimensional vector combined by the frequency $\omega$, amplitude $\beta$ and phase $\sigma$ of each segment.

To estimate the parameters $k$, $m_k$, $s_k$ and $\theta_k$, a reversible-jump MCMC (Markov chain Monte Carlo) algorithm [36] is applied. The reversible-jump MCMC is an extended version of standard MCMC that provides a simulation of the posterior distribution listed above on spaces of varying dimensions [36]. Thus, the simulation is possible even if the number of parameters in the model is unknown. This algorithm iterates between the segment model move and the change point model move according to its basic structure.

Before applying the AutoNOM, the maximal number of change points $k$ needs to be determined. Although the fitting results will be better as the value of $k$ increases, we want to avoid setting the $k$ too large, because when the value of $k$ is large, minor irregular fluctuations in the time series can falsely identify the existence of change points.

We use the mean average percentage error (MAPE) as a measure index to select the AutoNOM model with different input k values. MAPE is a method used to calculate the accuracy of curve fitting and is calculated as follows:

$$MAPE = \frac{1}{n} \sum_{i=1}^{n} \frac{|p_i - a_i|}{a_i} \times 100\%$$ 

(3)

where $a_i$ is the actual value, $p_i$ is the estimated value on minute $i$ and $n$ is the number of minutes for which the data is used.

## 4    Experiment

We use three different datasets in our experiment and process them in the following way:

- Dataset 1 - A synthetic dataset with known periods of 24, 36, and 48 hours (Figure 3). The sequences are comprised of different sinusoidal signals with a predefined frequency. We also add 3dB white noise to the dataset to simulate real conditions.
- Dataset 2 - A real-world open-source dataset [37] containing 70 consecutive days of heart rate and skin temperature collected in the one-minute interval as visualized in Figure 4. The values assumed to be missing at random (MAR) account for 7.63% of the whole dataset. We use the simple moving average (SMA) to impute the missing values [38]. SMA replaces the missing values by averaging the non-missing values within a rolling window without weights.
- Dataset 3 - A real-world dataset collected from Empatica E4 wristband, a medicalgrade physiological monitoring device [39], for over two weeks. The E4 device monitors the blood volume pulse (BVP), the electrodermal activity (EDA), the heart rate, and skin temperature in real-time with a sampling rate of 64Hz, 4Hz, 1Hz, and 4Hz respectively. We apply the same imputation method for processing data as in the second dataset.

### 4.1    Periodicity Detection

As shown in Table 1, all three methods can detect the periods of the synthetic dataset accurately with an error range between 0.09 and 0.36 hours. Among the three methods, the average error of FFT is the smallest, but the difference between FFT and the other two methods is small. These results verify the reliability of the three periodicity detection algorithms.

For the second dataset with a 70-day heart rate and temperature, the periodograms outputted by FFT and Chi-square are shown in Figures 5 and 6. In the Fourier periodogram, the period according to the dominant frequencies are around 24 and 12 hours. As for the Chi-square periodogram, the most significant periods for heart rate and skin temperature are 167.80 hours and 72.04 h, respectively. However, 168 h and 72 h are both multiples of 12 and 24 h, and the Chi-square periodogram also shows significant oscillations with a period of 24 hours. CyHMMs only infer a period of almost 24 hours for both time series (23.94 h and 24.01 h). Therefore, we can confirm that both heart rate and skin temperature have 24 h rhythms, which is consistent with the circadian cycle. Table 2 summarizes the detected results by the three methods.

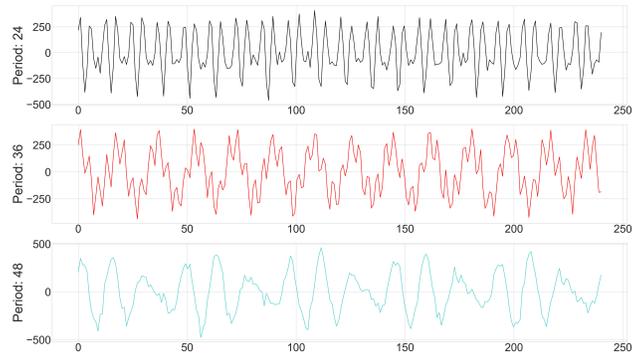

Fig.3: Visual inspection of the synthetic dataset

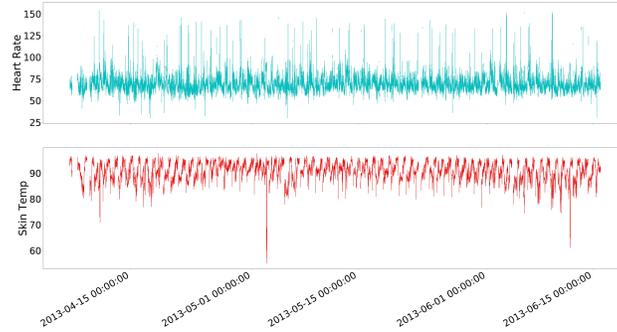

Fig.4: Visual inspection of the open-source dataset with heart rate and skin temperature data over 70 days

| Synthetic Period | Algorithm | Estimated | Error |
|---|---|---|---|
| 24 | FFT | 23.91 | 0.09 |
| | Chi | 23.89 | 0.11 |
| | CyHMMs | 23.64 | 0.36 |
| 36 | FFT | 35.72 | 0.28 |
| | Chi | 35.81 | 0.19 |
| | CyHMMs | 35.69 | 0.31 |
| 48 | FFT | 47.84 | 0.16 |
| | Chi | 47.79 | 0.21 |
| | CyHMMs | 47.87 | 0.13 |

Table 1: The performance of three period estimation algorithms on a synthetic dataset. The results from the synthetic dataset verify the reliability of the three periodicity detection algorithms.

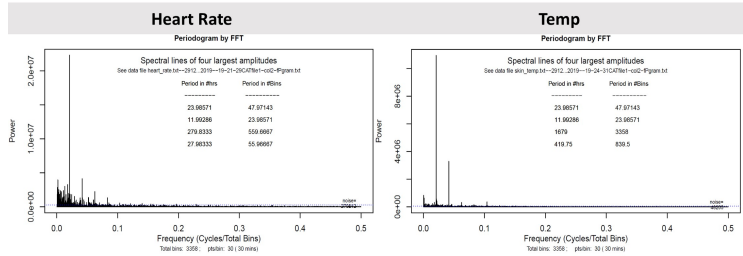

Fig.5: Fourier peridogram derived from heart rate, and skin temperature for the second dataset. The dashed line indicates the 0.05 level of significance for the periodogram. Dominant frequency correspond to 24 and 12 h for both heart rate and skin temperature.

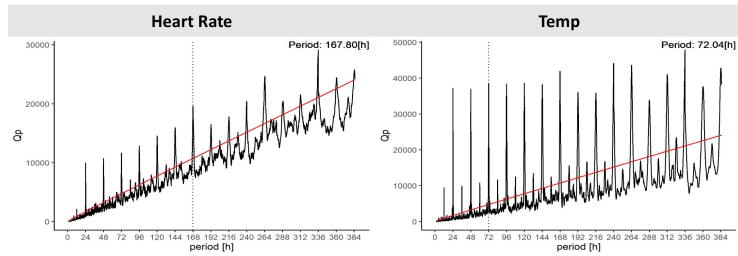

Fig.6: Chi-square periodograms derived from heart rate, and skin temperature for the second dataset. Red straight lines indicate the significance level of $p = 0.01$. Both heart rate and skin temperature in Chi-square periodogram exhibits a 24 h oscillation.

Compared to dataset 2, the E4 dataset contains new types of data, namely BVP and EDA, and the data is more fine-grained and clean. In terms of heart rate and skin temperature, FFT and Chi-square periodogram can display a period of 24 h, but the detected period is less significant than dataset 2, which may be a result of shorter time series (about 2 weeks only) and fine-grained high-frequency sampling rates that result in more fluctuations in the data.

In contrast to heart rate and skin temperature, the BVP and EDA show less significant periodic patterns. In Fig 7, BVP has a 24 h period, and EDA is detected to have a 25.6 h period more significant than the 24 h. In Fig 8, the Chi-square periodogram recognizes that the period of BVP is 23.95 h, which is close to the result derived from the FFT. The rhythm period of EDA is 50.12 h, but in the Chi-square periodogram of EDA, there also exists a peak $Q_P$ value above the significant level at around 25 hours, and this value corresponds to the 25.6 h in the FFT periodogram. The detected period of EDA by the CyHMMs is also 24.85 hours. As shown in Table 3, the results from the

| Physiological Data | Algorithm | Detected Periods |
| --- | --- | --- |
| | FFT | 23.98, 11.99, 279.83, 27.98 |
| Heart Rate | Chi | 167.80, 24.00, 12.00 |

|  | CyHMMs 23.94 |
|---|---|
| FFT | 23.98, 11.99, 1679, 419.75 |
| Skin Temperature Chi 72.04, 24.00, 12.00 | CyHMMs 24.01 |

Table 2: Periods of heart rate and skin temperature detected by FFT, Chi-square periodogram, and CyHMMs. Heart rate and skin temperature have obvious cycle characteristics of 24 h and 12 h. three methods reflect that EDA has a more extended period (around 25 hours) than the other three physiological signals. Besides 24 hours, from Fig 7, the period of 12 h is also significant in the Chi-square periodogram of EDA, heart rate and skin temperature, and the FFT periodogram of heart rate.

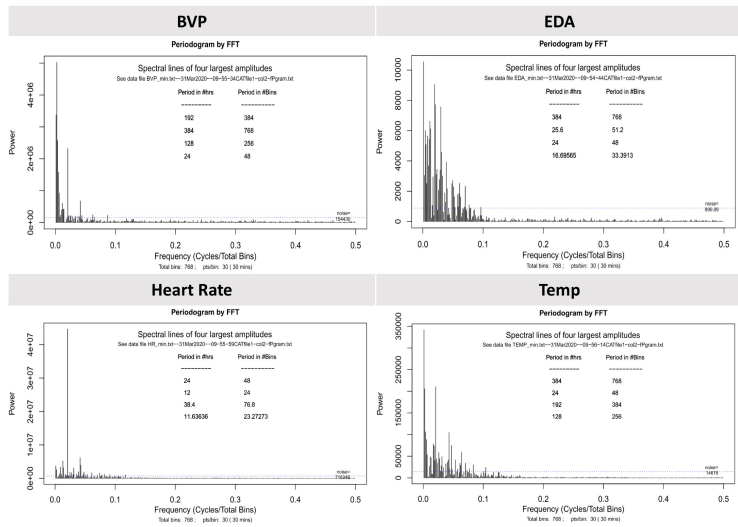

Fig.7: Fourier periodogram derived from Empatica E4 wristbant dataset. The dashed line indicates the 0.05 level of significance for the periodogram. Dominant frequency in BVP, heart rate and skin temperature corresponds to 24 h, or the integral multiple of 24 h. Period corresponding to main frequency in EDA is 25.6 h instead of 24 h. (384 is the total length of the dataset)

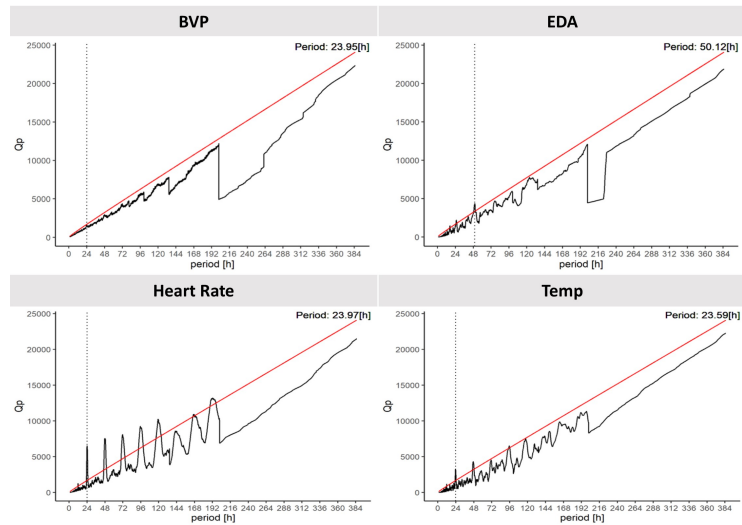

Fig.8: Chi-square periodograms derived from Empatica E4 wristband dataset. Red straight lines indicate the significance level of $p = 0.01$. Detected period in BVP is 23.95 h. Detected period in EDA is 50.12, and there is a a peak in the periodogram at around 25h. For heart rate and skin temperature, the periodograms display a 24 h oscillation.

| Physiological Data | Algorithm | Detected Periods |
|---|---|---|
| | FFT | 192.00, 384.00, 128.00, 24.00 |
| BVP | Chi | 23.95 |
| | CyHMMs | 23.97 |
| | FFT | 384.00, 25.60, 24.00, 16.69 |
| EDA | Chi | 50.12, 12.00, 24.00 |
| | CyHMMs | 24.85 |
| | FFT | 24.00, 12.00, 38.40, 11.63 |
| Heart Rate | Chi | 23.97, 12.00 |
| | CyHMMs | 24.00 |
| | FFT | 384.00, 24.00, 192.00, 128.00 |
| Skin Temperature | Chi | 23.59, 12.00 |
| | CyHMM | 23.83 |

Table 3: Periods of BVP, EDA, heart rate, and skin temperature detected by FFT, Chisquare periodogram, and CyHMMs. BVP, heart rate, and skin temperature own 24 h cycle, but the periodicity is not as evident as that of heart rate and skin temperature. The EDA detected by the three algorithms is about one hour longer than that of BVP, heart, and skin temperature.

## 4.2    Changing Point Detection in Cycles

As shown in previous sections, the FFT, chi-square periodogram, and CyHMMs have been validated in dataset 1 (synthetic data), and they are all able to identify a 24-hour rhythm period for heart rate and skin temperature in datasets 2 and 3. We continue our experiment on all three datasets to further detect the inner cycle and change points with the AutoNOM method within each period.

Dataset 1 is generated using the combination of constant frequencies and noise, and the AutoNOM identify changing points by perceiving changes in frequency, so it is reasonable that no change points have been detected in Fig 9. From another perspective, the AutoNOM has a strong anti-noise ability when searching for change points.

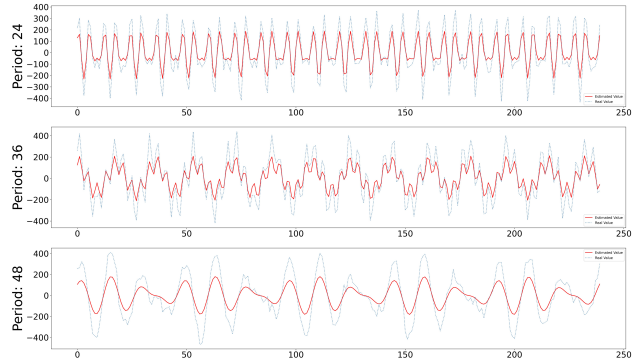

Fig.9: The analysis of dataset 1 using the AutoNOM. The blue lines are the real data, and the red lines depict the estimated values from the AutoNOM. No changing points have been detected by the AutoNOM, which is expected as there are no abrupt frequency changes in the synthetic dataset.

To demonstrate the performance of AutoNOM in detecting changing points in dataset 2, We choose one day from all 70 days in the dataset and use the AutoNOM to fit the one day's data. We choose $k$ equal to 4 as the optimal value based on an empirical observation: as shown in Table 4, when the value of $k$ changes from 3 to 4, the MAPE value drops significantly, and when the value of $k$ changes from 4 to 5, the change is minor. Figure 10 demonstrates similar distributions of the heart rate and the skin temperature change points where the two most frequent change points are located at around the 400th minute (6:40 am) and 1200th minute (8:00 pm) for both distributions. Illustrated in Figure 11, the skin temperature remains high at the start and end of each day, and it has a trough in the middle. As for the heart rate, it keeps low at the beginning of each day and goes through two combinations of up and down in turn and returns to a lower value at the end of the day. Between the two most frequent change points, the amplitudes and frequencies of the oscillation increases. Based on previous research, the body temperature of the circadian rhythm is under the control of the suprachiasmatic nucleus (SCN), which receives the input from photosensitive cells and synchronize body temperature and day and night alternation [40]. The temperature will reach its peak in the late afternoon and drop to its trough at the end of sleep [41]. In Figure 11, an abrupt decrease in skin temperature at the second changing point (about 7:00 am) and a stable increase between the second (about 7:00 am) and third changing point (about 5:00 pm) can be observed, which is consistent with previous studies. However, the difference is that the peak of skin temperature occurs in the late evening. Compared with body temperature, heart rate is less susceptible to the influence of the external environment [42,43]. The heart rate is more dependent on physical activity, and the heart rate during sleep is lower than during the wake time [44]. This view can explain why the heart rate in Figure 11 is lower on both sides of the day. The increase in heart rate in the two segments may be related to physical activity.

| | k=3 | k=4 | k=5 |
|---|---|---|---|
| Heart Rate | 4.54 | 3.29 | 3.14 |
| Skin Temp | 3.10 | 1.92 | 1.73 |

Table 4: The MAPE values of the AutoNOM method with different maximal number of change points $k$.

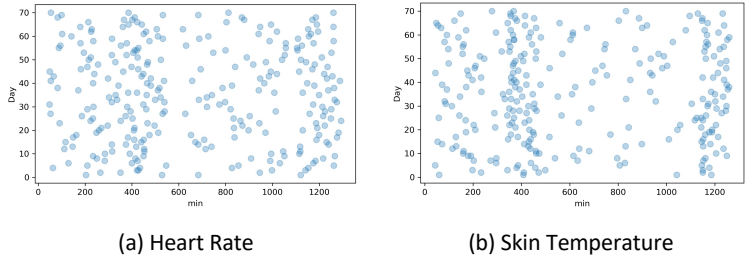

(a) Heart Rate                    (b) Skin Temperature

Fig.10: The distribution of change points in heart rate per day for 70 days. The x-axis represents the time of day in minutes. The change points are concentrated in the 400Th and 1200Th minutes of each day. Compared with the heart rate, the temperature has a more obvious tendency of the concentrated distribution. This change in heart rate and skin temperature is similar to the pattern of rest activity alternating between daytime activity and night rest.

We repeated the same process for dataset 3 and did the same analysis process on one random day in the data. The grey vertical lines in Figure 12 illustrates the changing points detected by the AutoNOM for heart rate, skin temperature, EDA, and BVP on the chosen day. The two common changing points for all four signals occur between the 400th minute (6:40 am) and 600th minute (10:00 am) and at around 1200th minute (8:00 pm), which are similar to the observation from dataset 2. In each segmentation between two changing points, the estimated curve output from the AutoNOM could fit the raw data well except for BVP. BVP reflects the relative change in blood volume caused by the heart contracting, so there will be many instantaneous and massive changes in BVP signals, which cause the AutoNOM not to work well. The daily trend of heart rate and skin temperature shown in Figure 11 and 12 are different, which could be caused by individual differences (e.g., lifestyle, daily schedule, and personality traits), climate, and even accuracy of wearable devices. We observe a seasonal effect in the skin temperature between the two datasets. The skin temperature in Figure 12 shows an apparent decrease between the second and third changing points, whereas the skin temperature in Figure 11 shows an increase at the same time. Checking the timing of data collection for these datasets, we found that dataset 2 was collected during spring, whereas dataset 3 was collected during the early winter. One interesting observation from dataset 3 is that there is one peak for heart rate during the first and second changing point, which is inconsistent with what we mentioned above. Peters et al. have found that the accelerated heart rate during sleep may be caused by uncomfortable sleeping posture [45]. When in an uncomfortable posture, the volume of intake oxygen will decrease, so the heart will increase the beat rate to demand oxygen supply, which is similar to what happens during strenuous exercise [46].

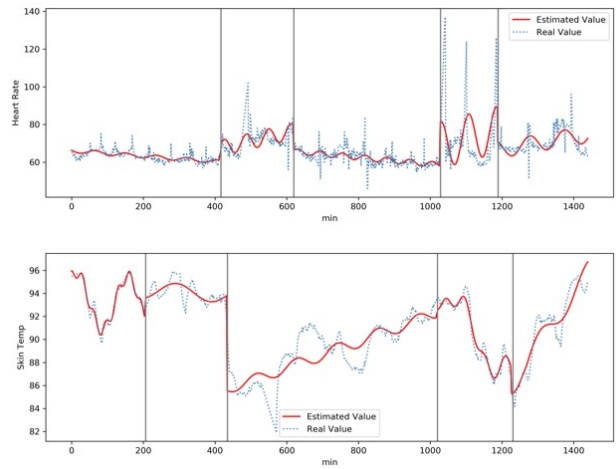

Fig.11: The analysis of heart rate and skin temperature in dataset 2 for one day using the AutoNOM. The blue lines are the real data, and the red lines depict the estimated values. The gray vertical lines show the estimated location of the change points. Three changing points occur at around 400th, 1000th, and 1200th in both heart rate and skin temperature. The heart rate increases in the second and fourth segments, while the remaining three segments keep low. The skin temperature maintains a high value at the beginning and end of the day, and there are a significant decrease and rebound in the middle.

Empatica E4 used in the dataset 3 is a medical-grade device that can collect accurate physiological data, but E4 can only work up to 40 hours and needs to be charged again. E4 cannot work when charging, and this process introduces missing values, which will cause unnecessary changing points. For example, the time points when E4 stops and starts working will be recognized as changing points. Due to the above uncontrollable factor, we do not provide a figure similar to Figure 10 for dataset 3.

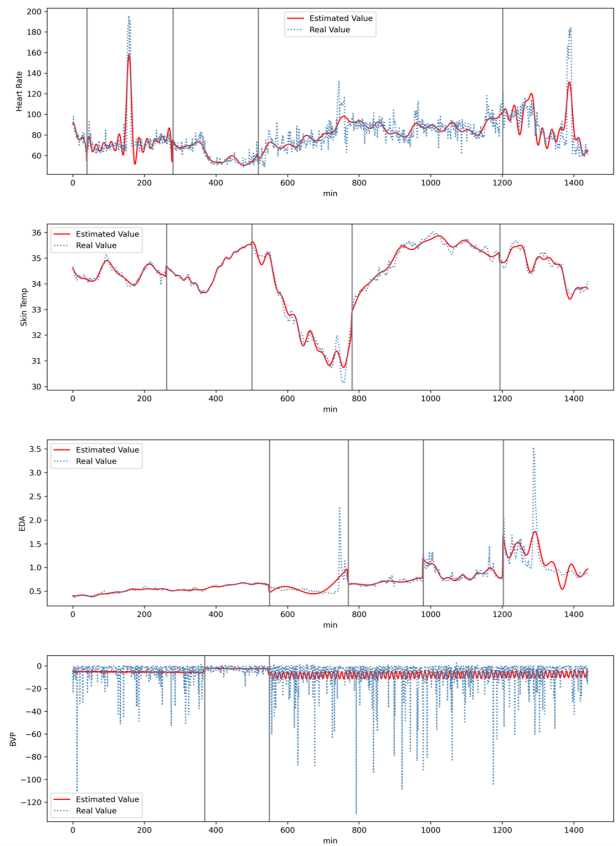

Fig.12: The analysis of heart rate, skin temperature, EDA, and BVP in dataset 3 for one day using the AutoNOM. The blue lines are the real data, and the red lines depict the estimated values from the AutoNOM.The gray vertical lines show the estimated location of the change points. The changing point between 400th and 600th minute is pretty close in the four signals. Heart rate, skin temperature, and EDA all have a changing point approximately at the 1200th minute. The heart rate fluctuates greatly at both ends, and the peak occurs at the beginning of the day. Similar to Figure 11, the trough of skin temperature appears in the middle. Except at the end of the day, EDA has little fluctuation. BVP declines in the second segment and has many instantaneous changes.

## 5    Implications and Conclusion

Our work explored the feasibility of designing a computational framework for automated discovery and modeling of human rhythms using data collected from consumer devices in the wild. We used three periodicity detection algorithms to examine their commonality in detecting periods in time series data without ground truth. The fact that those different methodologies identified the same periodic patterns in data implies the feasibility of incorporating those methods for periodicity detection in the framework to discover significant periods. FFT and Chi-Square periodograms output several significant periods compared to CyHMMs that output only the most significant one. Those periods indicate the existence of additional cycles in the same time series that are essential to discover and use in a rhythm-aware system for, e.g., planning. For the rhythm detection algorithms, the p-values of detected periods must exceed at a given significance level (e.g., the horizontal dash line in Figure 5 and 7, and the red line in Figure 6 and 8). If not, the input series will be assumed to have no periodicity. The periodicity detection methods on our three datasets identified multiple significant cycles in the biological data, including the presumed circadian 24 h cycle which confirms the existence of cyclic biobehaviors in humans.

We further explored the estimation of the change points in each period and observed the possibility of detecting inner cycles and fluctuations caused by external stimulus. This feature is dependent on the known period. As such, if the period detection algorithms cannot reach a consensus, our framework will be in a dilemma. Another shortcoming is that the maximum number of changing points $k$ needs to be tuned manually, which may mean a repetition of the process as the dataset is updated with new data.

While our exploratory study provides enough grounds for the development of a computational framework for modeling human rhythms from consumer-level wearable devices, we are curious to test these methods on behavioral data alone and in combination with physiological data. The modeling may reveal the causal relationship between behavioral and biological rhythms and whether they reflect people's mental and physical states. If so, we may be able to integrate our framework into health and well-being applications to provide interventions in the daily life of individuals according to their biological rhythms.